\documentclass[a4paper,unpublished]{quantumarticle}
\pdfoutput=1

\usepackage[colorlinks]{hyperref}
\usepackage[numbers,sort&compress]{natbib}
\usepackage[phfqit]{tchshortcuts}
\usepackage[
	print-unity-mantissa=false,
	exponent-product={\cdot},
	range-phrase=--
]{siunitx}
\usepackage{algorithm}
\usepackage{algpseudocodex}
\usepackage[capitalise,nameinlink]{cleveref}

\newcommand\oxfordaddress{Department of Materials,
University of Oxford,
Parks Road,
Oxford OX1 3PH,
United Kingdom}

\begin{document}

\title{Benchmarking Accuracy in an Emulated Memory Experiment}

\author{Tim Chan}
\email{timothy.chan@materials.ox.ac.uk}
\affiliation{\oxfordaddress}
\orcid{0000-0001-6187-7402}

\begin{abstract}
	This note proposes
	a simpler method to extract the logical error rate
	from an emulated surface code memory experiment.
\end{abstract}

\maketitle

The accuracy of a surface code decoder
is conventionally benchmarked on a classical computer by emulating
a memory experiment \cite{Gidney2022}.
The $\ket{\bar0}$ memory experiment,
for example:
prepares a logical qubit in $\ket{\bar0}$,
runs $n$ stabiliser cycles,
then measures it in the Z basis,
accounting for the corrections provided by the decoder
\cite[\S II.A]{Tan2023}.
The experiment
succeeds if the result matches the initial state $\ket{\bar0}$
and fails if it is $\ket{\bar1}$.

Emulation of such an experiment is often
done entirely on the decoding graph,
constructed from the detector error model \cite{Gidney2021a,Higgott2023}.
A \emph{logical bitflip} is a path of bitflipped edges
between opposite boundaries of the decoding graph.
Decoder accuracy is reported in terms of the \emph{logical error rate} $f(d)$,
which is the logical bitflip count per $d$ measurement rounds.
Our new method estimates $f(d)$ for decoders
that output a specific set of corrective edges to bitflip.

\section{Existing Method}
This method was first proposed in \cite[p~3]{Fowler2012}
but a fuller explanation can be found in \cite[\S A.3--4]{Tan2023}.
It works by repeating many times
the memory experiment of $n$ measurement rounds,
to estimate the experiment failure probability $f_n$.
This is done for various $n$
so that $f_n$ can be plotted against $n$;
a curve is then fitted to extract $f(d)$.

The advantage of this method is that
it sidesteps having to \emph{count} logical bitflips;
we instead need only determine
the \emph{parity} of said count $l$ in an experiment.
This is easily done by picking one of the two boundaries
and counting the number $l'$ of bitflipped edges it touches.
In general $l' \ne l$,
since paths of bitflipped edges between \emph{the same} boundary
contribute to $l'$ either 0 or 2 but to $l$ always 0.
However, this means their \emph{parities} are equal:
$l \bmod 2 =l' \bmod 2$.
Now,
let $k$ be the number of shots out of a total of $s$ for which $l$ is odd.
The experiment failure probability is estimated as
$\widehat{f_n} =k/s$
with a suitable confidence interval
like the \emph{Wilson score interval} \cite[\S 3.1.1]{Brown2001}.

The second part of this method
extracts $f(d)$ via a curve fit.
Intuitively,
$f_n$ should increase with $n$.
To model this dependence,
assume decoding each measurement round
leaves behind a logical bitflip independently with fixed probability $f(1)$.
Equivalently,
it erases
(forgets and replaces with a random $\ket{\bar 0}$ or $\ket{\bar 1}$)
the logical state with probability $2f(1)$.
The probability the state is \emph{not} erased after $\tau \in \mathbb Z_{\ge 0}$ rounds
equals the probability it is not erased after each intermediate round:
$1 -2f(\tau) =[1 -2f(1)]^\tau$.
So,
the experiment failure probability $f_n$ is given by
$1 -2f_n =\alpha[1-2f(1)]^n$
where the constant $\alpha$ accounts for errors from
preparation and measurement of the data qubits.
Plotting $\lg(1 -2f_n)$ against $n/d$
should thus yield a straight line
with gradient $\lg[1 -2f(d)]$.

In practice,
each memory experiment generates
\emph{multiple} Monte Carlo samples:
one for each value of $n$.
This is done by emulating the maximum number of measurement rounds,
then imagining if we stopped that experiment
at various shorter durations.
This reduces overall computation
but means samples are correlated.
The experiment must still be repeated many times
for each pair $(d, p)$ specifying the code distance and noise level,
respectively.

\section{New Method}
\begin{figure}[H]
	\centering
	\includegraphics[width=0.48\textwidth]{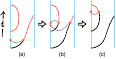}
	\caption{Sweeping upward through the decoding graph.
	The green horizontal line is the sweep line.
	The two blue vertical lines are boundaries.
	The black (red) curves are paths of
	encountered (unencountered) bitflipped edges.
	Each pair of red dots joined by a grey curve
	is an anyon pair.
	(a) An anyon pair is created.
	(b) Another pair is created.
	(c) The first pair spans opposite boundaries,
	so is recorded as a logical bitflip and removed.
	The second pair will later span between the same boundary,
	so will not contribute a logical bitflip.}
	\label{fig:new_method}
\end{figure}

A \emph{layer} of the decoding graph is the periodic unit subgraph
representing one measurement round.
An \emph{anyon pair} marks the endpoints of a path of bitflipped edges.
This method
sweeps through the decoding graph,
layer by layer,
and keeps track of a set of anyon pairs
for all encountered bitflipped edges.
\Cref{fig:new_method} illustrates an example.

Each newly encountered bitflipped edge in the current layer either
	updates the location of one anyon,
	creates a new anyon pair,
	or destroys an anyon pair.
When an anyon pair spans opposite boundaries
we record it as a logical bitflip and remove that pair.
When an anyon pair spans between the same boundary
we only remove that pair.
At the end of the memory experiment of $n$ measurement rounds
we should have zero anyon pairs left,
and a logical bitflip count $l$.
\Cref{sec:implementation} provides a concrete implementation for this.
The logical error rate is estimated as
$\widehat{f(d)} =ld/n$;
again the Wilson score interval is a suitable confidence interval.

\begin{figure}[H]
	\centering
	\includegraphics[width=0.48\textwidth]{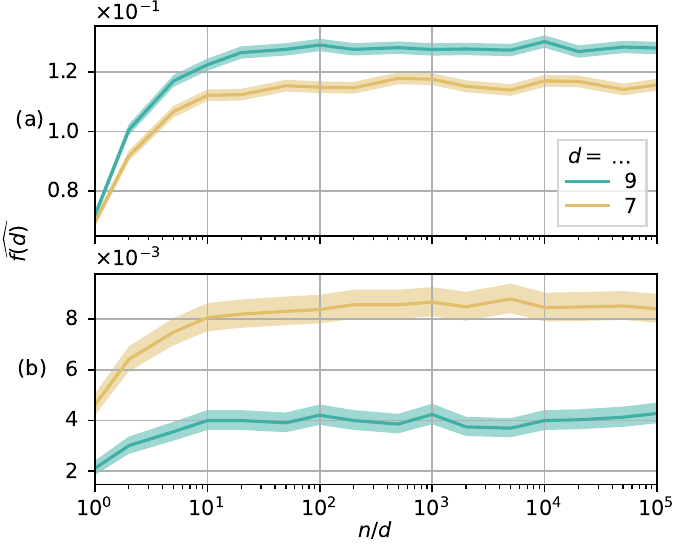}
	\caption{Estimated logical error rate of
	the Union--Find decoder \cite{Delfosse2020,Delfosse2021}
	adapted with the forward method \cite[\S VI.B]{Dennis2002},
	as a function of memory experiment duration.
	Each datapoint is the mean of $\num{1e5}$ lots of $d$ measurement rounds;
	shading shows the 95\% confidence region.
	(a) Noise level $p =\num{8e-3}$,
	which is above threshold;
	(b) $p =\num{4e-3}$,
	which is below threshold.}
	\label{fig:f_against_slenderness}
\end{figure}

This method simplifies the existing one
as we need only run one memory experiment for each $(d, p)$ pair.
The only requirement is that the experiment
must last long enough to make negligible
any transient effects at the start and end of the experiment.
\Cref{fig:f_against_slenderness} suggests
$\num{1e2}d$ measurement rounds is long enough.

\noindent
This method was introduced
to benchmark all the decoders in \cite{Chan2024};
the Python implementation used is
on GitHub at \cite{Chan2023a_quantum_bibstyle}.

\begin{acknowledgments}
I thank Simon Benjamin for useful discussions.
I acknowledge the use of
the University of Oxford Advanced Research Computing
(ARC)
facility~\cite{Richards2015_quantum_bibstyle} in carrying out this work
and specifically the facilities made available
from the EPSRC QCS Hub grant
(agreement No.\ EP/T001062/1).
I also acknowledge support from
an EPSRC DTP studentship
and two EPSRC projects:
RoaRQ (EP/W032635/1)
and SEEQA (EP/Y004655/1).
\end{acknowledgments}

\bibliographystyle{quantum}
\bibliography{tchbib}

\begin{thebibliography}{10}

\bibitem{Gidney2022}
Craig Gidney.
\newblock ``Stability experiments: The overlooked dual of memory experiments''.
\newblock \href{https://dx.doi.org/10.22331/q-2022-08-24-786}{{Quantum} {\bf 6}, 786}~(2022).

\bibitem{Tan2023}
Xinyu Tan, Fang Zhang, Rui Chao, Yaoyun Shi, and Jianxin Chen.
\newblock ``Scalable surface-code decoders with parallelization in time''.
\newblock \href{https://dx.doi.org/10.1103/PRXQuantum.4.040344}{PRX Quantum {\bf 4}, 040344}~(2023).

\bibitem{Gidney2021a}
Craig Gidney.
\newblock ``{Stim}: a fast stabilizer circuit simulator''.
\newblock \href{https://dx.doi.org/10.22331/q-2021-07-06-497}{{Quantum} {\bf 5}, 497}~(2021).

\bibitem{Higgott2023}
Oscar Higgott and Craig Gidney.
\newblock ``{Sparse} {Blossom}: correcting a million errors per core second with minimum-weight matching''~(2023).
\newblock  \href{http://arxiv.org/abs/2303.15933}{arXiv:2303.15933}.

\bibitem{Fowler2012}
Austin~G. Fowler, Adam~C. Whiteside, and Lloyd C.~L. Hollenberg.
\newblock ``Towards practical classical processing for the surface code''.
\newblock \href{https://dx.doi.org/10.1103/PhysRevLett.108.180501}{Physical Review Letters {\bf 108}, 180501}~(2012).

\bibitem{Brown2001}
Lawrence~D. Brown, T.~Tony Cai, and Anirban DasGupta.
\newblock ``Interval estimation for a binomial proportion''.
\newblock \href{https://dx.doi.org/10.1214/ss/1009213286}{Statistical Science {\bf 16}, 101--133}~(2001).

\bibitem{Delfosse2020}
Nicolas Delfosse and Gilles Z\'emor.
\newblock ``Linear-time maximum likelihood decoding of surface codes over the quantum erasure channel''.
\newblock \href{https://dx.doi.org/10.1103/PhysRevResearch.2.033042}{Physical Review Research {\bf 2}, 033042}~(2020).

\bibitem{Delfosse2021}
Nicolas Delfosse and Naomi~H. Nickerson.
\newblock ``Almost-linear time decoding algorithm for topological codes''.
\newblock \href{https://dx.doi.org/10.22331/q-2021-12-02-595}{{Quantum} {\bf 5}, 595}~(2021).

\bibitem{Dennis2002}
Eric Dennis, Alexei Kitaev, Andrew Landahl, and John Preskill.
\newblock ``Topological quantum memory''.
\newblock \href{https://dx.doi.org/10.1063/1.1499754}{Journal of Mathematical Physics {\bf 43}, 4452--4505}~(2002).

\bibitem{Chan2024}
Tim Chan.
\newblock ``Snowflake: A distributed streaming decoder''~(2024).
\newblock  \href{http://arxiv.org/abs/2406.01701}{arXiv:2406.01701}.

\bibitem{Chan2023a_quantum_bibstyle}
Tim Chan~(2023).
\newblock  code:~\href{https://github.com/timchan0/localuf}{timchan0/localuf}.

\bibitem{Richards2015_quantum_bibstyle}
Andrew Richards.
\newblock ``\href{https://doi.org/10.5281/zenodo.22558}{University of Oxford Advanced Research Computing}''.
\newblock ~(2015).

\end{thebibliography}

\appendix

\section{Implementation}
\label{sec:implementation}
To track the set of anyon pairs,
we can use a bidirectional map which we call \texttt{pairs}
e.g.\ if the set is $`{uv, wx}$,
then $\texttt{pairs}[u] =v$ and $\texttt{pairs}[v] =u$,
and similarly for $wx$.
Any pair $uv$ can be
added to the set with \texttt{pairs}.add($uv$),
and removed from it with either
\texttt{pairs}.remove($u$) or
\texttt{pairs}.remove($v$).

\Cref{alg:update_failure_count} summarises our new method.
The procedure \textsc{Load}
updates \texttt{pairs} with a new bitflipped edge,
simply ensuring each anyon is still
an endpoint of a path of bitflipped edges.

\begin{algorithm}[H]
\caption{Count logical bitflips in a memory experiment of $n$ measurement rounds.}
\label{alg:update_failure_count}
\begin{algorithmic}
	\State $\texttt{pairs} \gets$ empty bidirectional map
	\State $l \gets 0$
	\Comment{Initialise logical bitflip count.}
	\For{$k =1, \dots, n$}
		\For{each bitflipped edge $e$ in $k^\th$ layer}
			\State \Call{Load}{$\texttt{pairs}, e$}
		\EndFor
		\State $\texttt{new\_pairs} \gets$ empty bidirectional map
		\ForAll{$uv \in \texttt{pairs}$}
			\If{$uv$ spans opposite boundaries}
				\State $l \gets l +1$
			\ElsIf{$uv$ not on the same boundary}
				\State \Call{Load}{$\texttt{new\_pairs}, uv$}
			\EndIf
		\EndFor
		\State $\texttt{pairs} \gets \texttt{new\_pairs}$
	\EndFor
	\State \Output $l$
	\State
	\Procedure{Load}{$\texttt{pairs}, uv$}
		\If{$u \in \texttt{pairs}$}
			\State $w \gets\texttt{pairs}[u]$
			\State \texttt{pairs}.remove($u$)
			\If{$v \in \texttt{pairs}$}
				\State $x \gets\texttt{pairs}[v]$
				\State \texttt{pairs}.remove($v$)
				\State \texttt{pairs}.add($wx$)
			\ElsIf{$v \ne w$}
				\State \texttt{pairs}.add($vw$)
			\EndIf
		\ElsIf{$v \in \texttt{pairs}$}
			\State $x \gets \texttt{pairs}[v]$
			\State \texttt{pairs}.remove($v$)
			\State \texttt{pairs}.add($ux$)
		\Else
			\State \texttt{pairs}.add($uv$)
		\EndIf
	\EndProcedure
\end{algorithmic}
\end{algorithm}

\end{document}